\documentstyle[aps,twocolumn]{revtex}
\input{psfig}

\begin{document}

\def\hm{\ \rm {\it h}^{-1} Mpc}

\title{Throwing pebbles in the primordial pond}

\author{Carlo Baccigalupi}
\address{
INFN and Dipartimento di Fisica, Universit\`a di Ferrara, 
Via del Paradiso 12, 44100 Ferrara, Italy;\\
bacci@oarhp1.rm.astro.it
}

\maketitle
\begin{abstract} 
We consider anisotropies in the Cosmic Microwave Background 
(CMB) generated by spatially limited seeds; these objects 
could correspond to relics of high energy symmetry 
breaking in the early universe. 
It is shown how the CMB perturbation propagate beyond the 
size of the seed in the form of waves traveling with the CMB 
sound velocity. Moreover, these waves are the substantial 
part of the signal, both for polarization and temperature. 
The explanation of this phenomenology in terms of the CMB 
equations is given. 

Observationally, this effect is threefold promising. First, 
it enlarges the signal from a seed intersecting the last 
scattering surface to the scale of the CMB sound horizon 
at decoupling, that is roughly one degree in the sky. 
Second, it offers cross correlation possibilities between 
the polarization and temperature signals. Third, it allows 
to unambiguously distinguish these structures 
from point-like astrophysical sources. 
\end{abstract}

\section{The pebbles}

There is now a lot of interest in the connection between high 
energy physics and cosmology. It is motivated by the possibility 
that processes not reproducible here on the Earth actually 
occurred in the early universe. For this reason, a lot 
of work is currently in progress to predict in detail 
the traces that such processes could have left, in order 
to recognize them and gain insight into physics that is still 
unknown, or only theoretically approached. The unknown sector 
of physics extends from the energy scales presently explored by 
accelerators, described successfully by the standard model 
of the known elementary particles, up to the scales of more 
fundamental theories, perhaps supersymmetry and 
supergravity; such regimes, as thought in the early 
universe, should have taken place at temperatures $T$ 
(in energy units) in the interval 
\begin{equation}
10^{2}{\rm GeV}\le T\le 10^{19} {\rm GeV}\ {\rm or\ more}\ .
\label{vhep}
\end{equation}
\noindent
According to our hypotheses, two main classes of phenomena 
took place in the early universe: an era of accelerated expansion, 
the inflation, and the breaking of high energy symmetries, see 
\cite{LR}. 

The first process should leave traces in the form of 
Gaussian and scale-invariant density fluctuations \cite{MFB}; 
this visually corresponds to a completely disordered distribution 
of hills and wells in the density field, covering all the 
scales. 

The second process leaves completely different traces: 
spatially limited structures, like topological defects 
\cite{V} or bubbles made both of true and false vacuum \cite{CDLW}. 
At the present status of the theoretical knowledge, 
their production may occur with or without inflation. 
Models able to produce such structures both during 
and after inflation have been studied \cite{BGV,Y,KLS}. 
In order to be observable, the first case is most interesting, 
since the size of the structure is stretched by the stage of 
inflation after their formation, up to cosmologically interesting 
scales of tens of comoving Mpc or more. 

As well as the Gaussian fluctuations, these structures may 
be considered as seeds for the CMB perturbations. 
In the recent past, they have been thought as 
candidates for the structure formation process with preliminary 
discouraging results \cite{BR}, even if the numerical simulations 
and the models to explore are far to be exhausted; 
unfortunately, we do not have a good theory to predict 
their exact properties and abundance. The only 
sure thing is that the detection of at least one of them 
would be the first observational evidence of the existence 
of high energy symmetries. So the analysis here regards the signal 
from each single seed, without requiring neither that 
they dominate the structure formation process, nor 
that their signature is present on the whole sky CMB 
power spectrum. 

These seeds may also be thought to possess some 
spatial symmetries, both because appropriate and 
because the problem becomes simpler. 
Spherical and cylindrical symmetries are 
particularly simple and appropriate for bubbles, monopoles 
and strings, also forming loops \cite{BR,V}; 
also they allow to write simple and 
suitable formulas for the CMB perturbations; 
we refer to \cite{B} for a more quantitative and 
detailed exposition of these aspects. 

In this work we point out the characteristic signature of 
these structures on the CMB, in direct connection with 
the forthcoming whole sky CMB experiments \cite{PLANCKMAP}. 
As we shall see, their spatial shape combined with the 
undulatory properties of the CMB physics mix and produce 
their unambiguous signs. 

\section{The pond}

We begin with some necessary technical detail, but we hope to 
finish with physically simple and intuitive results. 

In linear theory, and assuming a Friedmann Robertson 
Walker (FRW) background, the equations driving the energy density 
perturbation and the peculiar motion of photons can be obtained 
from the linearized Einstein equations \cite{BKS}. 
Perturbations may be classified as 
scalar, vector and tensor with respect to spatial rotations; 
bubbles or topological defects are essentially fluctuations 
in the energy density composed by matter, radiation 
as well as scalar fields, 
therefore the case of interest here is the first one. 
The linearization implies a gauge freedom with 
respect to infinitesimal frame transformations; 
we choose the Newtonian gauge which physically corresponds 
to observers at rest with respect to the comoving expansion 
and experiencing the latter isotropic \cite{BKS,MB}. 

Perturbations in the CMB photons are coupled to the 
fluctuations of the other constituents of the cosmic 
energy density. In particular, Thomson scattering between baryons 
and photons induces polarization perturbations in the CMB, being 
an anisotropic process \cite{K}. At early times, the fluid is so 
dense that the photons free path $\dot{\tau}^{-1}$ vanishes; 
it is small with respect to the Hubble horizon $H^{-1}$ 
and the perturbation wavelength $1/k$ \cite{MB}. 
Therefore, the CMB equations may be expanded in powers of 
$k/\dot{\tau}$ and $H/\dot{\tau}$. In practice, the first 
order terms become important at decoupling, when the photons 
free path suddenly increases. 
One can consider CMB photons traveling on a direction 
$\hat{n}$ in the spacetime point $x\equiv (\eta ,\vec{x})$, 
where $\eta$ is the conformal time defined in terms of the ordinary 
time $t$ and of the scale factor $a$ by $d\eta =dt /a$. 
CMB temperature and polarization perturbations are expanded 
into spherical harmonics describing the dependence 
on $\hat{n}$. This treatment was firstly used in \cite{MB} 
and recently expanded to include non-flat geometries and 
non-scalar perturbations \cite{HSWZ}. 
For each Fourier mode, computations are performed in the 
$\hat{k}-$frame, where the wavevector $\hat{k}$ is the polar 
axis for the angular expansion; the fixed laboratory frame 
is instead indicated as the $lab-$frame; 
this distinction is particularly important for the 
perturbations considered here \cite{B}. 

To fix the ideas, before decoupling the CMB dynamics may be 
considered at the zeroth order in the Thomson scattering terms. 
Thus the equations for the energy density fluctuations $\delta$
and peculiar motion $v$ of photons are easily gained by the 
linearized conservation equations and have the simple 
following form: 
$$
\dot{\delta}=-{4k\over 3}v -4\dot{\Phi}
\ \ ,\ \ 
\dot{v}+{\dot{a}\over a}{3\rho_{b}\over 4\rho}
\left(1+{3\rho_{b}\over 4\rho}\right)^{-1}\cdot v=
$$
\begin{equation}
={k\over 4}
\left(1+{3\rho_{b}\over 4\rho}\right)^{-1}\cdot\delta +k\Psi\ \ .
\label{rho-v}
\end{equation}
$\Phi$ and $\Psi$ are the two scalar metric perturbations accounting 
for fluctuations from all the fluid species \cite{BKS}, and $a,\dot{a}$ 
are the cosmic scale factor and its derivative with respect to the 
conformal time. The terms containing the unperturbed baryon density 
$\rho_{b}$ are a residual of the coupling between photons and 
baryons, at the zeroth order in the Thomson scattering. 
Also we point out that $\delta ,v$ are simply linked to the 
monopole and the dipole of the CMB temperature fluctuation 
dependence on the photon propagation direction 
$\hat{n}$ \cite{MB,HSWZ}: 
\begin{equation}
\left({\delta T\over T}\right)_{0}={1\over 4}\delta\ \ ,\ \ 
\left({\delta T\over T}\right)_{1}=v\ \ .
\label{sb}
\end{equation}
Equations (\ref{rho-v}) may be put together in the following 
most simple form: 
$$
\ddot{\delta}+{\dot{a}\over a}{3\rho_{b}\over 4\rho}
\left(1+{3\rho_{b}\over 4\rho}\right)^{-1}\cdot\dot{\delta}+
{k^{2}\over 3}\left(1+{3\rho_{b}\over 4\rho}\right)^{-1}
\cdot\delta=
$$
\begin{equation}
=-{4k^{2}\over 3}\Psi -{\dot{a}\over a}{3\rho_{b}\over 4\rho}
\left(1+{3\rho_{b}\over 4\rho}\right)^{-1}
\cdot\dot{\Phi}-\ddot{\Phi}\ \ .
\label{rhov}
\end{equation}
This is a wave equation with friction of cosmological origin 
(proportional to $\dot{a}/a$) and forcing, gravitational 
terms at the second member. Focus on the friction term. Its 
effect is simple. Much before the horizon crossing the following 
conditions are satisfied \cite{BKS}:
the second term of the first member dominates over the third one, 
the gravitational potentials are constant and $\rho\gg\rho_{b}$; 
therefore the solution is trivially $\delta =-4\Psi$ 
and no friction effect exists at all. At the horizon crossing 
the last term of the first member becomes important, and $\delta$ 
starts to oscillate, making $\dot{\delta}$ not vanishing. Thus the 
friction becomes active and damps the oscillations. Now we 
come to the central arguments of the present work. 

Differently from analogous problems 
in cosmology, the friction term is multiplied by 
$\rho_{b}/\rho$. In most cosmological models the baryon content 
is of the order of percent with respect to the dark matter component, 
because of the severe constraints from nucleosynthesis \cite{O}. 
Therefore it is evident that 
\begin{equation}
10^{-2}\le {\rho_{b}\over\rho}\le 10^{-1}
\label{new} 
\end{equation} 
between equivalence and decoupling, 
reducing substantially the friction term for the 
oscillations occurring at these epochs. 

For the perturbations considered here, 
the forcing terms in (\ref{rhov}) are active on spatially 
limited regions, occupied by the seed. The arguments above show 
that outside the seed the equation for $\delta$ is 
\begin{equation}
\ddot{\delta}+{k^{2}\over 3}\delta\simeq 0\ \ .
\label{hc}
\end{equation}
This says that the oscillations occurring at the 
horizon crossing are brought {\it outside} the seed with the 
CMB sound velocity. They shall reach the sound horizon at the time 
considered, given by
\begin{equation}
h_{s}(\eta)\simeq\int_{0}^{\eta}{1\over\sqrt{3}}d\eta  '=
{\eta\over \sqrt{3}}\ ,
\label{hs}
\end{equation}
and it can be shown that it corresponds roughly 
to one degree in the sky; we remark that 
$\delta$ waves mean $\delta T/T$ waves from (\ref{sb}), 
and this means polarization waves, since polarization and 
temperature perturbations are tightly coupled \cite{K,HSWZ}. 

The consequences of the exposed arguments are straightforward. 
Consider a spatially limited seed intersecting the last scattering 
surface. Its CMB signal is made of waves extending approximatively 
over one degree in the sky. This is extremely interesting for the 
future whole sky, high resolution observations \cite{PLANCKMAP}. 
These same waves are a unambiguous proof of the fact that really 
the seed was generated well before decoupling, simply because 
any other astrophysical source does not produce them because it formed 
after decoupling, appearing point-like. 
These concepts might be observationally decisive 
if our theoretical thinking is quite right and high energy 
symmetry breaking traces are really present in the nearby universe. 

In the next section we give a numerical example of the 
realization of the phenomenology exposed here in the context 
of the standard cosmological scenario. We refer to \cite{B} for 
all the computational and formal details. 

\section{The waves}

We considered spherical and infinitely long cylindrical 
energy concentrations in a background formed by 
cold dark matter $(\Omega_{CDM}=.95)$, baryons 
$(\Omega_{b}=.05)$, photons and massless neutrinos, 
assumed distributed initially adiabatically. 
At any time, and for this kind of sources, the expressions 
of the CMB temperature and polarization perturbations may 
be written in simple and intuitive forms \cite{B}. 
We give here the expressions for spherical sources. 

{\bf Temperature}. The temperature CMB perturbation 
at any spacetime point $(\eta ,\vec{r})$ for a spherical source 
is given by \cite{B}: 
\begin{equation}
{\delta T\over T}=
\sum_{l\ge 0}P_{l}(\hat{n}\cdot\hat{r})
\int_{0}^{\infty}{k^{2}dk\over 2\pi^{2}}
\left({\delta T\over T}\right)_{l}(k,\eta )
j_{l}(kr)\ .
\label{dtt}
\end{equation}
$P_{l}$ and $j_{l}$ are the Legendre polynomials and 
the fractional order Bessel function respectively. 
Note how the spatial directions $\hat{n}$ and $\hat{r}$, 
describing geometrically the problem, are {\it outside} 
the Fourier integral. In spite of the virtually infinite 
series, already the $l=0$ term is the substantial 
component of the signal, as we show below. 

{\bf Polarization}. It is described by the 
Stokes parameters $Q$ and $U$ on the plane perpendicular 
to the photon propagation direction $\hat{n}$. 
As for the temperature case, each Fourier mode of 
the polarization perturbations admit an expansion in 
spherical harmonics with coefficients $Q_{l}$ and $U_{l}$. 
Remembering that these quantities are defined in the 
$\hat{k}-$frame, the problem may be further simplified since 
$U_{l}=0$ for scalar perturbations. 
Also we mention that as a distinction with respect 
to the temperature case, the {\it tensor} spherical harmonics 
are now required, accounting for the tensor properties of 
the polarization; for this reason, $Q_{l}$ and 
$U_{l}$ are defined for $l\ge 2$ \cite{HSWZ}. 
The polarization CMB perturbation 
at any spacetime point $(\eta ,\vec{r})$ in presence 
of a spherical source is given by \cite{B}: 
\begin{equation}
Q=\cos (2\phi_{\hat{r}})\cdot{\cal I}
\ \ ,\ \ U=-\sin (2\phi_{\hat{r}})\cdot{\cal I}\ ,
\label{p}
\end{equation}
$$
{\cal I}=\sum_{l\ge 2}\sqrt{(l-2)!\over (l+2)!}
P_{l}^{2}(\hat{n}\cdot\hat{r})
\int_{0}^{\infty}{k^{2}dk\over 2\pi^{2}}Q_{l}(k,\eta )j_{l}(kr)\ . 
$$
$Q$ and $U$ are essentially identical except for the geometrical 
dependence on $\phi_{\hat{r}}$, that is the angular coordinate 
of the projection of $\vec{r}$ on the plane perpendicular 
to $\hat{n}$, as seen in the $lab-$frame: if the latter is 
chosen so that $\phi_{\hat{r}}=0$ we have $U=0$. Note that now 
the {\it second order} Legendre polynomials compare into the sum. 
They are meaningful since prevent photons traveling radially 
($\hat{n}\cdot\hat{r}=\pm 1$) to be polarized; this is correct since 
in that case no preferred axis exists for polarization. 
As for the temperature case, the substantial component of the 
perturbation is given by the first term of the sum, $l=2$. 

Note that these relations are {\it independent} on the nature of 
the seed, which is encoded in the $(\delta T/T)_{l},Q_{l},U_{l}$ 
coefficients and could be made 
of matter and radiation as well as scalar field, that has the only 
restriction in its sphericity. 
Each coefficient $(\delta T/T)_{l},Q_{l},U_{l}$ obeys 
motion equations, and together with the linearized Einstein ones 
the whole system may be evolved in time, 
see \cite{B} and references therein. 

\begin{figure}
\psfig{figure=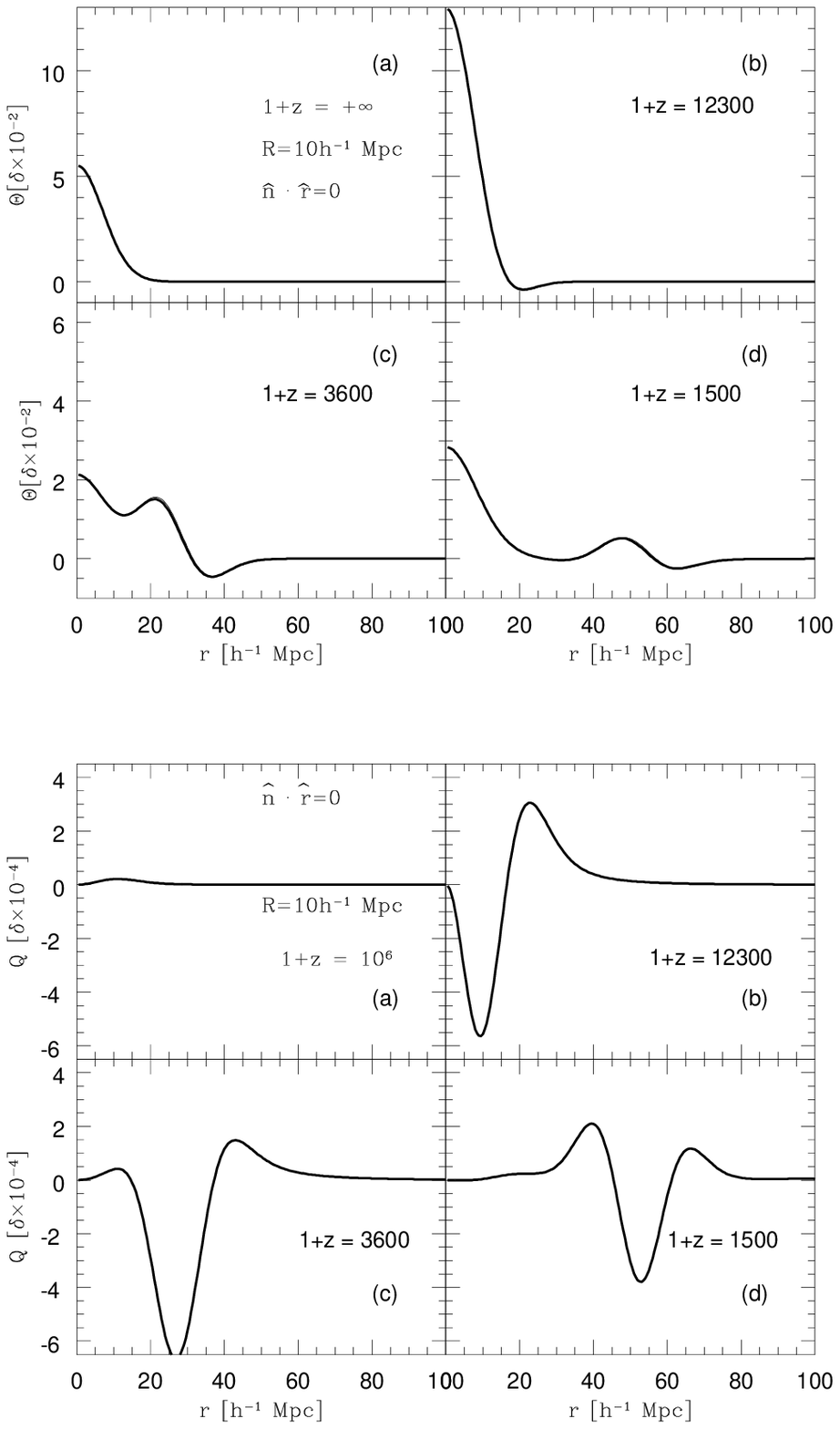,height=7.in,width=4.in}
\caption{CMB perturbation waves from a spherical seed at different 
times and as a function of the radial distance. The seed has 
the indicated comoving radius. Waves form at the horizon crossing 
and travel outward, both for temperature (up, indicated as $\Theta$) 
and polarization (down, indicated as $Q$). 
They are well visible just before decoupling and 
are the unique sign of the previous history of the seed itself.}
\label{lf1}
\end{figure}

\begin{figure}
\psfig{figure=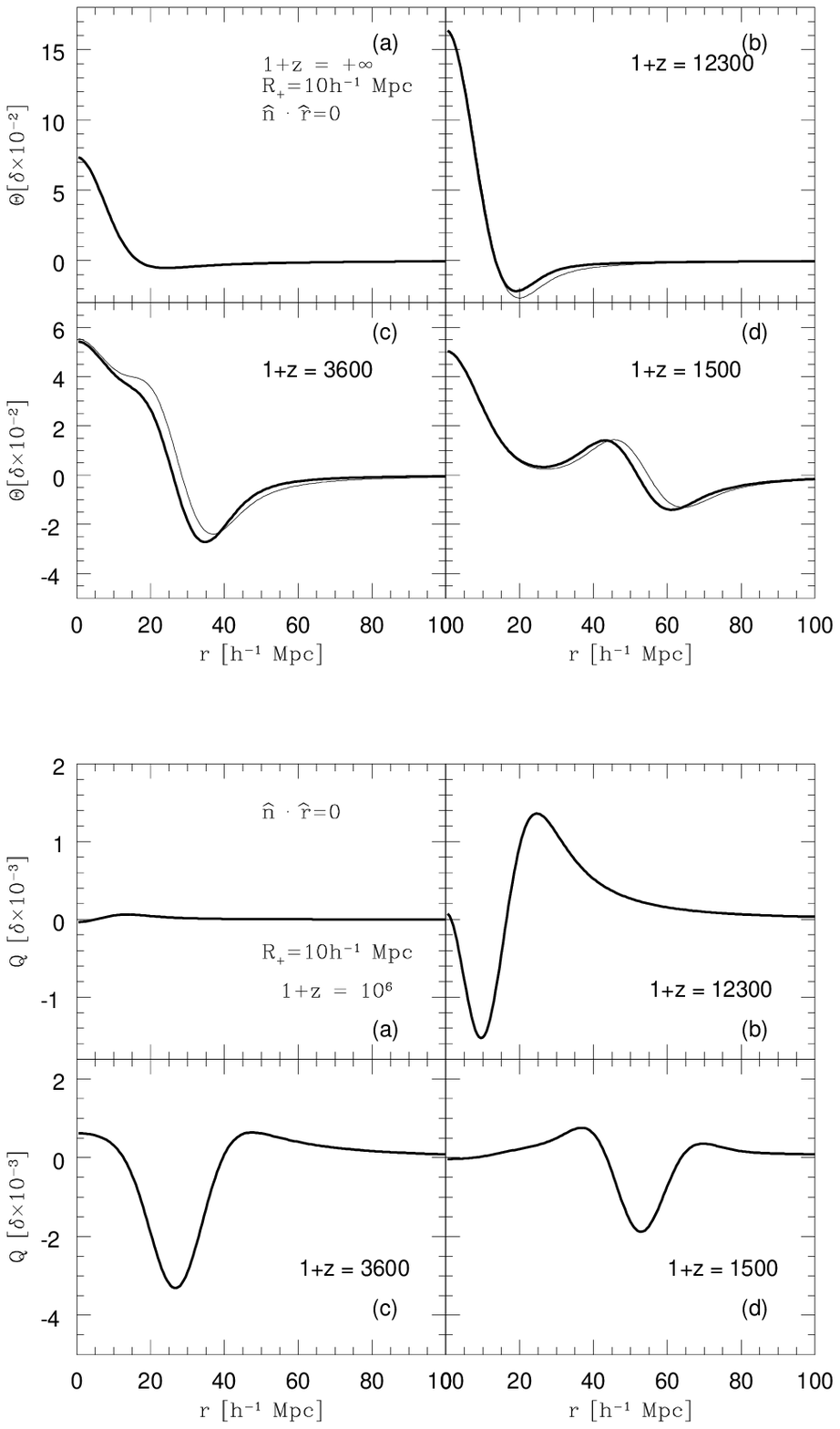,height=7in,width=4.in}
\caption{CMB perturbation waves from a cylindrical seed at different 
times and as a function of the distance from the symmetry axis. 
Waves form at the horizon crossing and travel outward like in 
figure \ref{lf1}.}
\label{lf2}
\end{figure}

Figures \ref{lf1} and \ref{lf2} show the time evolution 
of the CMB temperature and polarization perturbations. 
In each panel the perturbation profile is shown as 
a function of the distance from the seed center and symmetry 
axis respectively for the spherical and cylindrical case. 
The perturbations are normalized with the density 
contrast at decoupling $\delta$, taken at the center for the 
spherical seed and on the symmetry axis for the cylindrical one. 
At the horizon crossing, CMB waves form and travel 
outward, just like the waves from a pebble thrown in a pond. 

The upper panels regards the temperature signals, 
while the lower ones shows the polarization amplitude. 
The thin lines represents the signal only from the 
monopole term in (\ref{dtt}), while the thick ones 
contains all the contributions; this shows how the 
monopole is the dominant term. As indicated, 
photons travel perpendicularly to the 
radial distance $\hat{r}$ for the spherical case and 
on the plane perpendicular to the symmetry axis for 
the cylindrical case. 
The symmetries of the seed allow to choose 
the $lab-$frame axes so that the polarization 
is given by a pure $Q$ term \cite{B}. In the 
spherical case they are simply parallel and 
perpendicular to the plane formed by 
$\hat{n}$ and the radial direction $\hat{r}$ 
in the scattering point; in the cylindrical case 
they are the symmetry axis $\hat{z}$ and the 
direction perpendicular to $\hat{n}$ and $\hat{z}$. 

Several interesting comments may be made on these 
graphs. 

The waves at any time occupy the 
position of the CMB sound horizon (\ref{hs}), and 
have roughly the same amplitude of the perturbation 
inside the seed. Really, in the polarization case, 
they are the very dominant component of the signal. 
Also a marked correlation is evident between the 
polarization and temperature waves. 

The symmetries of the seeds constrain the signal. 
Photons traveling radially in a spherical seed must 
be not polarized since this problem is symmetric 
with respect to rotations around the propagation 
direction $\hat{n}$. In the cylindrical case 
the symmetry axis itself is a preferred direction 
and polarization perturbations may affect 
photons traveling away from the symmetry axis. 

Again we remark that this undulatory behavior 
of the CMB perturbations occurs for seeds 
existing since the beginning, $\eta =0$; 
no waves at all arise from sources 
formed after decoupling. 

\begin{figure}
\psfig{figure=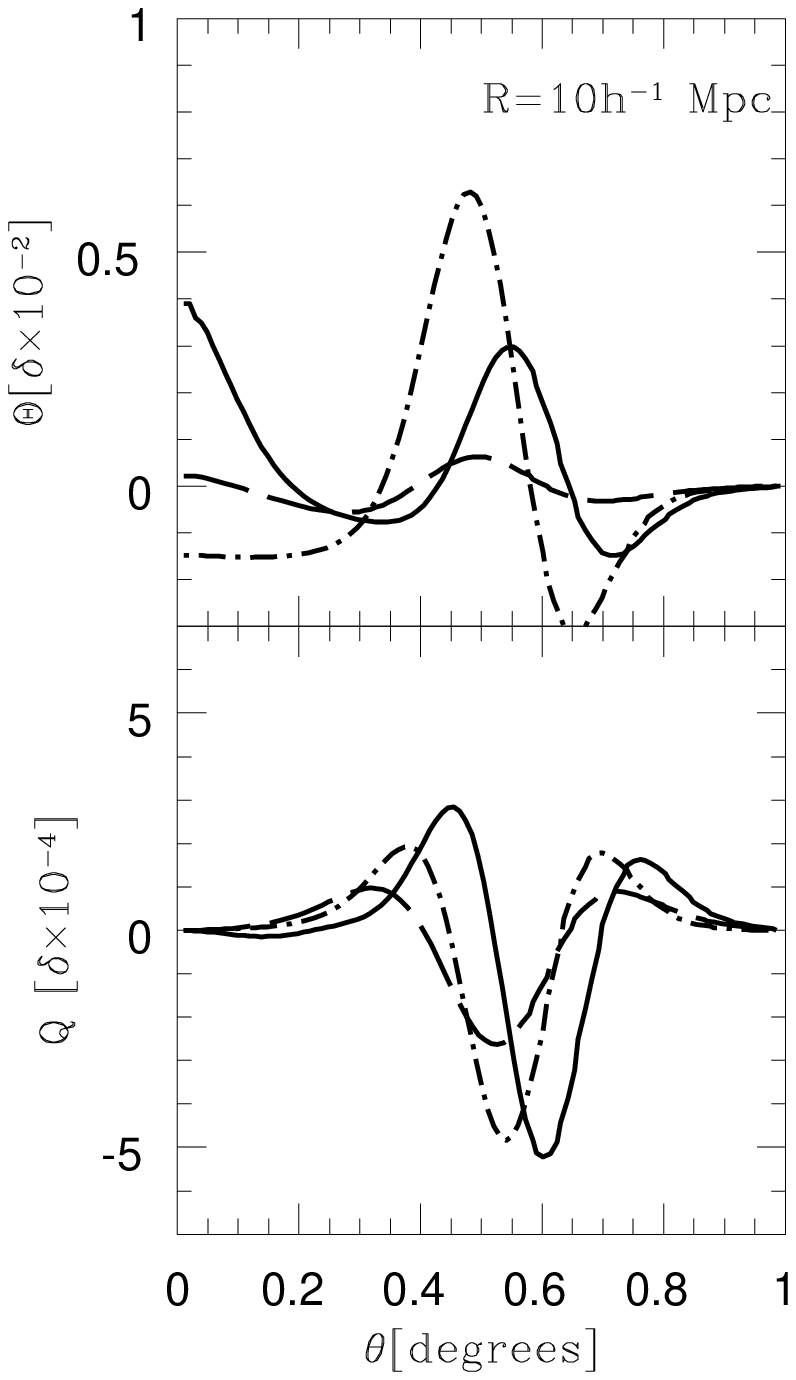,height=4in,width=4.in}
\caption{CMB anisotropy from a spherical seed as a function of 
the angular distance $\theta$ from the central direction. 
The waves as well as the central temperature spot evident 
in figure \ref{lf1} have been photographed by the 
decoupling photons and are visible to us.}
\label{lf3}
\end{figure}

These figures represent the CMB perturbations 
around spatially limited seeds at different times. 
Now we show in the spherical case
the real simulation of the CMB anisotropies, 
in order to show how the exposed behavior is 
maintained \cite{B}. Figure \ref{lf3} shows the CMB 
anisotropy from the spherical seed if it intersects 
the last scattering surface, as a function of the 
angle $\theta$ from the center. The Sachs-Wolfe effect 
regarding the zone physically occupied by the seed, 
$\theta\le 10'$ has been included. The CMB 
perturbation waves have been photographed by the 
decoupling photons. Of course the signal is a 
function of the relative disposition of the seed 
with respect to the last scattering surface. 
Different cases have been displayed: 
the solid line represents the case in which the seed 
center lies exactly on the last scattering surface; 
the dashed and dotted dashed lines shows the signals 
if the seed lies 30$h^{-1}$ Mpc within or outside our 
Hubble sphere. In particular, if the distance between seed 
and last scattering surface is much larger then a 
CMB sound horizon at decoupling, roughly 100$h^{-1}$ 
Mpc, the seed is visible only through the 
integrated Sachs-Wolfe effect if it lies within 
our Hubble sphere; in this case however the CMB 
waves could not be detected, even if distinctive 
signals would arise \cite{BBS}. 

The amplitude of the signal is roughly as expected for 
linear perturbations with density contrast $\delta$ and 
size $L$ smaller than $H^{-1}$ at decoupling: 
$\delta T/T\simeq \delta\cdot (LH)^{2}$ and a few percent 
of this for the polarization amplitude. 

Of course a lot of work has to be done to predict in detail 
the signs from any relic of the early universe; at the moment 
such prediction exists only for the temperature signals 
from inflationary bubbles \cite{Bi}. 

The final general statement that we make here is the following. 
If the forthcoming high resolution CMB maps 
should contain perturbed spots surrounded 
by anisotropy waves reaching the distance of the CMB 
sound horizon at decoupling, then we could conclude that 
the processes that generated such signals belong to the unknown 
very high energy sector of physics. This would open new 
possibilities for testing fundamental physical theories. 

\acknowledgements

This work was written at the SISSA/ISAS institute. 
The author is warmly grateful to the Astrophysics Sector 
for the kind hospitality.


\onecolumn

\end{document}